\documentclass[a4paper,11pt]{article}
\usepackage{graphicx} 
\usepackage{amsmath}
\usepackage{amssymb}
\usepackage{amsfonts}
\usepackage{geometry}
\usepackage{microtype}
\usepackage{babel}
\usepackage{amsthm}
\usepackage[hypertexnames=false]{hyperref}
\usepackage{comment}
\usepackage[utf8]{inputenc}
\usepackage{xcolor}

\hypersetup{
    colorlinks=true,   	
    linkcolor=red,      
    citecolor = [rgb]{0 0.7 0},   	
    filecolor=magenta, 	
    urlcolor=blue
}

\newcommand{\RL}{{\mathbb R}}

\newcommand{\IND}{{\mathbb I}}
\newcommand{\BBP}{{\mathbb P}}

\def\ba{\begin{align}}
\def\ea{\end{align}}
\def\ban{\begin{align*}}
\def\ean{\end{align*}}

\def\be{\begin{eqnarray}}
\def\ee{\end{eqnarray}}
\def\ben{\begin{eqnarray*}}
\def\een{\end{eqnarray*}}

\def\bqq{\begin{equation}}
\def\eqq{\end{equation}}
\def\bqqn{\begin{equation*}}
\def\eqqn{\end{equation*}}

\def\sq{$\Box$}

\def\qed{\ifmmode\sq\else{\unskip\nobreak\hfil
\penalty50\hskip1em\null\nobreak\hfil\sq
\parfillskip=0pt\finalhyphendemerits=0\endgraf}\fi\par\medbreak}

\newsavebox{\junk}
\savebox{\junk}[1.6mm]{\hbox{$|\!|\!|$}}

\def\det{{\mathop{\rm det}}}

\def\til={{\widetilde =}}

\def\clC{{\cal C}}

\def\clN{{\cal N}}

 \def\eq#1/{(\ref{#1})}

\newtheorem{theorem}{Theorem}[section]
\newtheorem{corollary}[theorem]{Corollary}

\newtheorem{lemma}[theorem]{Lemma}

\def\eq#1/{(\ref{e:#1})}

\def\bdes{\begin{description}}
\def\edes{\end{description}}

\def\notes#1{}

\definecolor{mag}{rgb}{0.7,0,0.3}
\definecolor{dgreen}{rgb}{0.1,0.5,0.1}
\definecolor{dred}{rgb}{.8,0,0}
\definecolor{gray}{rgb}{.8,.8,.8}
\definecolor{brown}{rgb}{0.6451,0.3706,0.1745}

\newcommand{\R}{\mathbb{R}}
\DeclareMathOperator{\TV}{TV}

\newcommand{\E}{\mathbb E}

\title{$L^1$ stability of the 
entropy power inequality in ${\mathbb R}^d$\\
without moment assumptions}

\author{%
Norman El Fartass\thanks{%
D\'{e}partement de Math\'{e}matiques,
\'{E}cole Normale Sup\'{e}rieure Paris-Saclay,
91190 Gif-sur-Yvette, France.
Email:
\texttt{norman.el\_fartass@ens-paris-saclay.fr}.
}
\and 
Lampros Gavalakis\thanks{%
Statistical Laboratory, Centre for Mathematical Sciences, 
University of Cambridge, Wilberforce Road, Cambridge CB3 0WB, U.K.
Email: 
\texttt{lg560@cam.ac.uk}.
}
\and 
Ioannis Kontoyiannis\thanks{%
Statistical Laboratory, Centre for Mathematical Sciences, 
University of Cambridge, Wilberforce Road, Cambridge CB3 0WB, U.K.
Email: 
\texttt{yiannis@maths.cam.ac.uk}.
}
}
\date{\today}

\begin{document}

\maketitle

\begin{abstract}
We establish qualitative stability 
in $L^1$ and Ruzsa distance 
for Shannon's entropy power inequality 
in $\mathbb R^d$,
without any moment assumptions. 
The proof uses 
the optimal transport argument by Rioul (2017),
majorization bounds of density rearrangements,
and the stability of the sharp 
form of Young's inequality
due to Christ (2019).
Along the way we also extend recent
one-dimensional
results on 
de Bruijn's identity, 
the continuity of differential entropy,
and the
equality case of the entropy power
inequality
to random vectors
of any dimension.
\end{abstract}

\section{Introduction and main results}

One of the equivalent formulations of Shannon's~\cite{shannon:48}
celebrated 
entropy power inequality (EPI) states that, for any pair of 
independent random vectors $X, Y$ in $\mathbb{R}^d$ and 
any $\lambda \in (0, 1)$,
$$
h(\sqrt{\lambda}X + \sqrt{1-\lambda}Y) 
\geq \lambda h(X) + (1-\lambda)h(Y), 
$$
where $h(X)$ denotes the differential entropy of a random 
vector $X$, given by
\[
h(X) = -\int_{\mathbb{R}^d} f(x) \log f(x) \, dx,
\]
when $X$ has density $f$ and the integral exists,
and by $h(X)=-\infty$ otherwise.
The first proof of the EPI
was given by 
Stam in 1959~\cite{stam:59} and made use of the 
de Bruijn identity,
\begin{equation}
\frac{d}{dt} h(X + \sqrt{t}Z) 
= \frac{1}{2} I(X + \sqrt{t}Z), \quad t > 0, \label{eq:deBruijn}
\end{equation}
where $Z \sim \mathcal{N}(0, I_d)$ is a standard Gaussian random 
vector in $\mathbb{R}^d$ independent of $X$,
and $I(X)$ denotes 
the Fisher information of a random vector $X$,
defined as 
$$
I(X) = \int_{\mathbb{R}^d} \frac{\|\nabla f(x)\|_2^2}{f(x)} \, dx,
$$
whenever $X$ has a weakly differentiable
density $f$, and as $I(X) = +\infty$ otherwise~\cite{bobkov:22}.

Our main goals are
$(i)$~to show that, without
any additional assumptions, there is equality in
the EPI if and only if
$X$ and $Y$ are Gaussian with the same covariance
matrix (Theorem~\ref{dEPITh});
and $(ii)$~to establish a qualitative
stability result for the EPI
in $L^1$ under minimal assumptions 
(Theorem~\ref{prop:tv}).

\subsection{Equality conditions}

Although
Stam only made use of de Bruijn's identity in dimension one 
(proving the EPI in dimension~$d$ using an inductive argument),
the multidimensional identity~\eqref{eq:deBruijn} 
has numerous applications~\cite{dembo-cover-thomas:91} and it
also appears in 
the work of Bakry and {\'E}mery~\cite{bakry:book}
as a measure of entropy dissipation 
along the heat flow. 

Using~\eqref{eq:deBruijn}, the EPI is derived 
as a consequence of 
Stam's inequality, which is 
the convolution inequality 
\begin{equation} \label{fisherinfoineq}
I(\sqrt{\lambda}X+\sqrt{1-\lambda}Y) 
\leq \lambda I(X) + (1-\lambda)I(Y), 
\end{equation}
for any $\lambda \in (0,1)$, with equality,
when $I(X)$ and $I(Y)$ are finite, if and only if $X$ 
and $Y$ are Gaussian with the same covariance 
matrix~\cite[Theorem~7]{carlen:91b}. 
The rigorous justification of the exchange of differentiation 
and expectation in de Bruijn's identity~\eqref{eq:deBruijn} 
in dimension $d=1$ was first established by Barron in 
1984~\cite{barron:cltTR} under the assumption of finite variance. 
Recently, de Bruijn's identity was shown to hold 
under the necessary and sufficient condition that $h(X+Y) < \infty$ for 
some random variable $Y$ with finite $h(Y)$ independent of $X$ in 
dimension $1$~\cite{EPI:26toappear,gavalakis-stability:ISIT26}.

The following two results extend a key
entropy continuity property 
and de Bruijn's identity to any dimension $d\geq 1$.
Their proofs, given in Section~\ref{episec},
are straightforward generalizations of 
the one-dimensional 
arguments~\cite{EPI:26toappear,gavalakis-stability:ISIT26}.

\begin{theorem}[Entropy continuity]
\label{thm:continuity_gauss}
Let $X$ be a random vector in $\mathbb{R}^d$ with finite differential entropy, for which there exists an independent random vector $Y$ with finite differential entropy such that $h(X+Y) < \infty$. Then, if $Z \sim \mathcal{N}(0, I_d)$ is independent of X,
\begin{align*}\lim_{t\downarrow 0} h(X + \sqrt{t}Z) = h(X).\end{align*}
\end{theorem}

\begin{theorem}[De Bruijn's identity]
\label{thm:deBruijn}
Let $X$ be a random vector in $\mathbb{R}^d$ with finite differential entropy, for which there exists an independent random vector $Y$ with finite differential entropy such that $h(X+Y) < \infty$. Then, if $Z \sim \mathcal{N}(0, I_d)$ is independent of $X$,
\[
\frac{d}{dt} h(X + \sqrt{t}Z) = \frac{1}{2} I(X + \sqrt{t}Z), \quad t > 0.
\]
\end{theorem}

Although the EPI is known to hold in complete generality,
the case of equality has only recently been characterized
in the case $d=1$
without additional moment conditions or regularity assumptions
of the underlying densities; see~\cite{EPI:26toappear}
for a detailed historical account.
As described in Section~\ref{episec},
Theorems~\ref{thm:continuity_gauss} and~\ref{thm:deBruijn} allow us 
to justify the steps in the classical proof of the EPI and thus 
to characterize the equality case when $d\geq 2$ in full generality:

\begin{theorem}[EPI] 
\label{dEPITh}
Let $X,Y$ be independent random vectors on $\R^d$. 
Then, for any $\lambda\in(0,1),$
$$
h(\sqrt{\lambda}X + \sqrt{1-\lambda}Y) \geq \lambda h(X) + (1-\lambda)h(Y).
$$
If $h(X)$ and $h(Y)$ exist and are finite, there is equality if and only if $X$ and $Y$ are Gaussian with the same (finite) covariance matrix. 
\end{theorem}

\subsection{Qualitative stability}
A qualitative stability result in terms of weak convergence
was shown in~\cite{EPI:26toappear,gavalakis-stability:ISIT26}
for the EPI in $\mathbb R$,
where an extensive discussion
of the literature around the stability of the EPI is also given. 
It is known~\cite{courtade:18} that there is no 
stability of the EPI in 
relative entropy and quadratic Wasserstein distance,
even under the assumption of finite second moments.

Our main result in this work,
Theorem~\ref{prop:tv} below,
shows that there is
qualitative stability in $L^1$, under minimal assumptions.

As in~\cite{EPI:26toappear}, for independent 
random vectors $X,Y$, we let the ``EPI deficit''
$\delta_{{\rm EPI},\lambda}(X,Y)$ be
$$
\delta_{\mathrm{EPI},\lambda}(X,Y):= 
h(\sqrt{\lambda}X + \sqrt{1-\lambda}Y) - \lambda h(X) - (1-\lambda)h(Y).
$$
For simplicity we write $\delta_{\mathrm{EPI}}(X,Y) 
:= \delta_{\mathrm{EPI},\frac12}(X,Y)$ 
and $\delta_{\mathrm{EPI}}(X) 
:= \delta_{\mathrm{EPI}}(X,X')$ when $X,X'$
are independent and identically distributed
(i.i.d.).
Also, 
we adopt the
common notation
$$
N(X)=\frac{e^{2\frac{h(X)}{d}}}{2\pi e},
$$
for the entropy power of $X$,
and we write
\(\phi_{m,\Sigma}\) for the Gaussian density
with mean $m$ and covariance matrix \(\Sigma\succ0\)
on $\RL^d$.
 
\begin{theorem}[Qualitative $L^1$ stability]
\label{prop:tv}
Suppose \(\{X_n\}_{n\geq1}\) and
\(\{Y_n\}_{n\geq1}\) are sequences of random vectors in \(\R^d\)
with corresponding densities \(\{f_n\}_{n\geq1}\) and
\(\{\widehat f_n\}_{n\geq1}\), 
such that, for each $n$,
$X_n$ is independent of $Y_n$ and 
$h(X_n)$ and $h(Y_n)$ are finite.
If
\[
  \delta_{\mathrm{EPI}}(X_n,Y_n)
  \rightarrow0,
\]
then
\[
  |h(X_n)-h(Y_n)|\rightarrow0.
\]
Moreover, writing $h_n = \sqrt{N(X_n)N(Y_n)}$,
\[
  \inf_{\substack{
    m,\widehat{m}\in\R^d\\
    \Sigma\succ0,\ \det(\Sigma)=1}}
	\Big[
    \left\|
      f_n-
      \phi_{m,h_n\Sigma}
    \right\|_1
	+
    \left\|
      \widehat f_n-
      \phi_{\widehat{m}, h_n\Sigma}
    \right\|_1
	\Big]
  \rightarrow0.
\]
\end{theorem}

In the discrete setting, the corresponding
``EPI deficit'' at $\lambda=1/2$ is often called
the {\em Ruzsa distance} $d_R(X,Y)$ 
between two independent discrete random variables
$X,Y$
taking values in the same additive group,
$$d_R(X,Y):=H(X-Y)-\frac12H(X) -\frac12 H(Y),$$
where $H(X)$ is the discrete Shannon entropy of $X$.
Stability estimates 
for the best lower bound on $d_R(X,Y)$
are tightly connected to sharp 
inverse sumset theorems in additive 
combinatorics~\cite{tao:10,green:25}. 
For example, the entropic inequality underlying
the proof of Marton's conjecture over \(\mathbb F_2^n\)~\cite{GGMT:23b} 
states that
there exists a subgroup \(H\subset\mathbb F_2^n\) such that
\[
d_R(X,U_H)+d_R(Y,U_H)\le 11\,d_R(X,Y),
\]
where \(U_H\) is uniform on \(H\).
A torsion-free analogue would assert that, for some $C<\infty$,
\begin{equation} \label{discreteG}
d_R(X,G)+d_R(Y,G)\le C\,d_R(X,Y),
\end{equation}
where \(G\) is an appropriate notion of a discrete Gaussian; 
see the discussion in \cite[Section~5]{raghavan:25}.

Since Gaussian random vectors are inherently continuous objects, 
one might be tempted to think that the continuous 
version of~\eqref{discreteG}
would be natural and easier to prove. Nevertheless, the obvious such 
analogue turns out to be false. For any two 
random vectors $X,Y$ in $\RL^d$, let
\bqq
d_R(X,Y):=h(X'-Y')-\frac{1}{2}h(X)-\frac{1}{2}h(Y),
\label{eq:RuzsaC}
\eqq
where $X',Y'$ are independent versions
of $X,Y$.
If \(U_N\) is uniform on \(\{0,\ldots,N-1\}\) and independent of
\(Z\sim\mathcal N(0,1)\), then \(X_N=N^2U_N+Z\) satisfies,
as $N\to\infty$,
\[
d_R(X_N,X_N) = O(1),
\]
yet for a Gaussian $G_N$ with the same entropy as $X_N$
it is not difficult to verify that
$$
d_R(X_N,G_N)
=\log N+O(1).
$$

It is a consequence of Theorem~\ref{prop:tv}
that, in the i.i.d.\ case, the continuous 
analogue of~\eqref{discreteG} {\em is} true in $\mathbb R^d$, provided that 
the ``doubling" $d_R(X,-X)$ 
is close enough 
to its minimum possible value of $\frac{d}{2}\log 2$:

\begin{corollary}\label{cor:bound}
For every \(C>1\), there exists
\(\varepsilon_0(C,d)>0\) such that every
\(\mathbb R^d\)-valued random vector \(X\) with finite differential
entropy satisfying
\[
  d_R(X,-X')\le\frac d2\log 2+\varepsilon_0(C,d),
\]
where $X'$ is an independent copy of $X$,
also satisfies
\[
  d_R(X,Z)\le C\,d_R(X,-X'),
\]
for a Gaussian \(Z\) with \(h(Z)=h(X)\).
\end{corollary}

Theorem~\ref{prop:tv} and Corollary~\ref{cor:bound} are proved 
in Section~\ref{stabilitySec}. The proof of Theorem~\ref{prop:tv} uses 
a modification of Rioul's optimal transport argument for the 
EPI~\cite{rioul:17} to obtain convergence of the logarithms 
of the Jacobian determinants of the
transport maps to zero in probability. This facilitates the choice 
of an appropriate conditional density, the spherically symmetric 
decreasing rearrangement of which is then shown to converge in entropy. 
This implies convergence in $L^2$ due to majorization,
and the
result follows by an 
application of a relatively recent 
stability result for the sharp form of Young's 
convolution inequality due to Christ~\cite{christ:19}. 

\section{The EPI}
\label{episec}

The proofs in this section are mostly direct generalizations of the 
corresponding proofs in~\cite{EPI:26toappear}.

Let $\mathcal{C}_{\rm BC}$ denote the
class of random vectors $X$ in $\mathbb{R}^d$ with finite differential 
entropy, and such that $h(X+Y) = +\infty$ for every random vector $Y$ 
independent of $X$ with $h(Y) > -\infty$. In one dimension,
Bobkov and Chistyakov~\cite{bobkov:15}
constructed an explicit example of a random variable $X$ 
showing that $\clC_{\rm BC}$ is nonempty.
The same argument generalizes to $\mathbb{R}^d$.
Taking $(X_1,\dots,X_d)$ with i.i.d.\ coordinates $X_i$ distributed as 
the one-dimensional example $X$, we have 
that $h(X_1,\dots,X_d) = d\,h(X_1)$ is finite 
while $h((X_1,\dots,X_d)+(Y_1,\dots,Y_d)) 
= d \cdot h(X_1+Y_1) = +\infty$,
for any vector $(Y_1,\ldots,Y_d)$ of
i.i.d.\ coordinates with $h(Y_1)>-\infty$. 
Using the submodularity for 
sums inequality~\cite{KM:14,KM:16}, 
we deduce 
that $\mathcal{C}_{\rm BC} \neq \emptyset$ in every dimension.

The following three results, stated without proof,
are straightforward generalizations of 
the corresponding one-dimensional results in~\cite{EPI:26toappear}.

\begin{lemma}
\label{lem:sec_moment_conv}
Let $X$ be a random vector in $\mathbb{R}^d$ with entropy $-\infty < h(X) < + \infty$ such that there exists a random vector $Y$ independent of $X$, with finite entropy and with $h(X+Y) < +\infty$. Then $h(X+Z) < + \infty $ for any Z independent of $X$ with finite second moment. 
\end{lemma}

\begin{theorem}
\label{thm:dichotomy}
Let $X$ be a random vector with $- \infty < h(X) < +\infty$. Then either:
\begin{enumerate}
    \item $X\in \clC_{\rm BC}$, \text{ or}
    \item $h(X+Y) < +\infty$ for all $Y \notin \clC_{\rm BC}$ with 
	finite entropy.
\end{enumerate}
\end{theorem}

\begin{lemma}
\label{lem:gauss_conditioned}
Let $X$ be a random vector in $\mathbb{R}^d$ with finite entropy, $Z$ a standard Gaussian vector independent of $X$, and $t_0 > 0$. Assume that $h(X + \sqrt{t_0}Z) < \infty$. Then:
\begin{align*}
\lim_{n \xrightarrow \infty} 
h\big(X + \sqrt{t_0}Z \,\,\big|\,\, \|X\|_2 \leq n\big) = h(X + \sqrt{t_0}Z).
\end{align*}
\end{lemma}

\begin{proof}[Proof of Theorem~\ref{thm:continuity_gauss}]
The result of the theorem follows from
Lemma~\ref{lem:sec_moment_conv},
Theorem~\ref{thm:dichotomy} and Lemma~\ref{lem:gauss_conditioned},
exactly as in the case of its
one-dimensional 
counterpart~\cite[Theorem~3]{EPI:26toappear}.
\end{proof}

Next we recall the classical fact that the 
convolution between the normal density and any other density 
satisfies the heat equation; see, e.g.,~\cite[Theorem~4.3]{folland:book}.

\begin{lemma}
\label{lem:lemma16}
Let $X$ be a continuous random vector in $\mathbb{R}^d$ 
with density $f$,
and let $g_t = f \ast \varphi_t$, 
where $\varphi_t:=\phi_{0,tI_d}$.
Then $g_t$ satisfies the heat equation, i.e., 
$$
\frac{\partial}{\partial t}g_t(y) = \frac{1}{2} \Delta_y \, g_t(y), \quad \textit{for every} \,\, t > 0, \, y \in \mathbb{R}^d,
$$
where $\Delta_y \, g_t(y) = \sum_{i=1}^d \frac{\partial^2}{\partial y_i^2}g_t(y).$
\end{lemma}

The next lemma follows from Lemma~\ref{lem:lemma16}
in exactly the same way as the proof 
of~\cite[Lemma~17]{EPI:26toappear} in
one dimension.

\begin{lemma}
\label{lem:lemma17}
Let $X$ be a random vector with density $f$, with
finite differential entropy, and for which there exists 
an independent random vector $Y$ with finite differential entropy 
itself such that $h(X+Y) < \infty$. 
Let $Z$ be a standard Gaussian vector independent of $X$.
Then, writing $g_t = f \ast \varphi_t$, we have:
$$
\frac{\partial}{\partial t}h(X + \sqrt{t}Z) 
= - \int \left( \frac{\partial}{\partial t}g_t(y) \right) 
\log g_t(y)dy, \quad t>0.
$$
\end{lemma}

The last lemma we need is the following
integration-by-parts
representation for the Fisher information
of a random vector perturbed by a Gaussian.

\begin{lemma}
\label{lem:lemma18}
Let $X \in \mathbb{R}^d$ be a random vector with density $f$, finite 
differential entropy, and for which there exists an independent random 
vector $Y$ with finite differential entropy itself such that $h(X+Y)$ 
is finite. 
Let $Z$ be a standard Gaussian vector independent of $X$.
Then,
$$
I(X + \sqrt{t}Z) = -\int_{\mathbb{R}^d} 
\left( \Delta_y\, g_t(y) \right) 
\log g_t(y)\, dy, \quad t > 0.
$$
\end{lemma}

\begin{proof}
Fix $t>0$ and write $g_t$ for the density of $X+\sqrt{t}Z$.
Lemmas~\ref{lem:lemma16} and~\ref{lem:lemma17}
imply that
$\left( \Delta_y\, g_t(y) \right) \log g_t(y)$
is integrable.
Let $F(s)=s\log s-s$, $s>0$.
Since $g_t$ is smooth and positive, 
\bqq
\Delta_y F(g_t(y))
=(\Delta_y g_t(y))\log g_t(y)
+\frac{\|\nabla g_t(y)\|_2^2}{g_t(y)}.
\label{eq:Deltay}
\eqq
Also, \(h(X+\sqrt t\,Z)\) is finite by 
Lemma~\ref{lem:sec_moment_conv}
and the boundedness of $g_t$.
Therefore,
$g_t\log g_t$ and $F(g_t)$ are both integrable.

Consider a family $\{\chi_R;R>0\}$ of smooth, compactly supported
cutoff functions on $\RL^d$, with 
$0\leq\chi_R\uparrow 1$ and $\|\Delta\chi_R\|_\infty=O(R^{-2})$,
as $R\to\infty$; see, e.g.,~\cite[Chapter~2]{lee:book}. 
Multiplying~\eqref{eq:Deltay} by $\chi_R$ and integrating,
integration by parts gives
$$\int\chi_R\Delta_y F(g_t)=\int F(g_t)\Delta_y\chi_R,$$
and, rearranging, 
\[
\int_{\mathbb R^d}\chi_R(y)\frac{\|\nabla g_t(y)\|_2^2}{g_t(y)}\,dy
=
\int_{\mathbb R^d}F(g_t(y))\Delta_y\chi_R(y)\,dy
-
\int_{\mathbb R^d}\chi_R(y)(\Delta_y g_t(y))\log g_t(y)\,dy.
\]
Taking $R\to\infty$,
the first integral on the right-hand side converges to zero
by bounded convergence, 
since its absolute value is at most 
$\|F(g_t)\|_1\|\Delta\chi_R\|_\infty$,
the second integral converges,
by dominated convergence, to 
\[
-\int_{\mathbb R^d}(\Delta_y g_t(y))\log g_t(y)\,dy,
\]
and the integral on the left-hand side converges
to $I(X+\sqrt{t}Z)$ by monotone convergence.
\end{proof}

\begin{proof}[Proof of Theorem~\ref{thm:deBruijn}]
    De Bruijn's identity follows by combining 
Lemmas~\ref{lem:lemma16},~\ref{lem:lemma17} and~\ref{lem:lemma18}.
\end{proof}

\begin{proof}[Proof of Theorem~\ref{dEPITh}]
Using the $d$-dimensional versions of 
de Bruijn's identity in Theorem~\ref{thm:deBruijn}
and of the entropy
continuity result in Theorem~\ref{thm:continuity_gauss},
Stam's one-dimensional argument as developed
in~\cite[Appendix]{EPI:26toappear} 
immediately extends to $\mathbb {R}^d$. 
The equality case characterization thus follows 
from the equality case in the Fisher 
information 
inequality~\eqref{fisherinfoineq}~\cite[Theorem~7]{carlen:91b}
combined with Cram{\'e}r's theorem~\cite{cramer:36}.
\end{proof}

\section{Stability} 
\label{stabilitySec}
This section is devoted to the proof of the qualitative stability result, 
Theorem~\ref{prop:tv}.

\subsection{Preliminaries}\label{preliminarysec}

For a pair random vectors $X,Y$ with densities $f_X,f_Y$ on $\RL^d$, 
respectively, their total variation distance is
$$
\TV(X,Y) = \frac12\|f_X-f_Y\|_1 = \frac12\int_{\mathbb R^d}|f_X(x)-f_Y(x)|\,dx.
$$
Recall that, if the differential entropies of $X$ and $Y$ 
are finite, the (continuous) Ruzsa distance~\cite{KM:14,KM:16} 
between $X$ and $Y$ was defined as in~\eqref{eq:RuzsaC},
in analogy to the 
discrete setting~\cite{tao:10,green:25}.
Recall also the definition of the EPI deficit 
$\delta_{\mathrm{EPI}}(X,Y)$ from the Introduction. 

We will use the following continuity statements: 

\begin{lemma}\label{lem:regularization}
Suppose \(h(X)\) and \(h(X+X')\) are finite. If \(G\) is an independent
standard Gaussian, then \(h(X+\sqrt t\,G)\) is finite for every \(t>0\),
and as $t\downarrow 0$ we have:
\begin{align*}
  h(X+\sqrt t\,G)&\rightarrow h(X),             
	\\
  \delta_{\mathrm{EPI}}(X+\sqrt t\,G)&\rightarrow\delta_{\mathrm{EPI}}(X),
	\\
  \TV(X+\sqrt t\,G,X)&\rightarrow0.              
\end{align*}
\end{lemma}

\begin{proof}
The first two statements are direct consequences
of Theorem~\ref{thm:continuity_gauss}.
The $L^1$ convergence in the third statement is a standard
fact about Gaussian kernels, 
see, e.g.,~\cite[Theorem~4.3]{folland:book}.
\end{proof}

 For a nonnegative integrable function \(f:\R^d\to[0,\infty)\),
its spherically symmetric decreasing 
rearrangement~\cite{lieb-loss:book,wang:14} \(f^\star\)
is the radially nonincreasing function, unique up to null sets,
such that
\[
  \bigl|\{x\in\R^d:f^\star(x)>t\}\bigr|
  =
  \bigl|\{x\in\R^d:f(x)>t\}\bigr|,
  \qquad t>0.
\]
Equivalently,
\[
  f^\star(x)
  =\int_0^\infty
    \IND\{f>t\}^\star(x)\,dt,
\]
where, for $A\subset \mathbb R^d,$ \(A^\star\) denotes the ball centered at the origin
with the same Lebesgue measure as \(A\).

Rearrangement preserves
\(L^p\) norms and entropy whenever the latter is 
finite~\cite{wang:14}. We will make use of the 
following estimate to pass from convergence of 
entropies to $L^2$ convergence.

\begin{lemma}\label{lem:majorization}
Let \(u,v\) be probability densities on \(\R^d\) with finite
entropies, and let \(u,v\leq M<\infty\). Suppose that \(v\) majorizes
\(u\), that is,
\[
  \int_{B_r}u^\star(x)\,dx
  \leq \int_{B_r}v^\star(x)\,dx,
\]
for all $r>0,$ where $B_r = \{x\in \mathbb R^d: \|x\|_2 <r\}$.
Then
\[
  0\leq \|v\|_2^2-\|u\|_2^2
  \leq 2M\bigl(h(u)-h(v)\bigr).
\]
\end{lemma}

\begin{proof}
By~\cite[Proposition~VII.3]{wang:14}, majorization implies
\begin{equation} \label{madprop}
  \int_{\R^d}\Psi(u(x))\,dx
  \leq \int_{\R^d}\Psi(v(x))\,dx
\end{equation}
for every convex function \(\Psi:[0,\infty)\to\R\) continuous
at zero with \(\Psi(0)=0\), provided both integrals are well defined.
Applying this with \(\Psi(t)=t^2\), we obtain the claimed lower bound,
$\|u\|_2^2\leq\|v\|_2^2$.

For the upper bound, define
\[
  \Psi(t)=
  \begin{cases}
    t\log t-\dfrac{t^2}{2M},&0<t\leq M,\\[1mm]
    M\log M-\dfrac M2+(t-M)\log M,&t>M,\\
    0,&t=0.
  \end{cases}
\]
This function is continuous at zero. It is convex since
\[
  \Psi''(t)=\frac1t-\frac1M\geq0
  \quad(0<t<M),
  \qquad
  \Psi'(M-)=\Psi'(M+)=\log M,
\]
and it is affine on \([M,\infty)\).
Moreover, since $u$ is bounded and has finite entropy
\[
  \int_{\R^d}|\Psi(u)|
  \leq \int_{\R^d}u|\log u|
       +\frac{\|u\|_2^2}{2M}<\infty,
\]
and similarly for \(v\). Applying~\eqref{madprop} again gives,
\[
  -h(u)-\frac{\|u\|_2^2}{2M}
  \leq
  -h(v)-\frac{\|v\|_2^2}{2M},
\]
which implies the desired upper bound.
\end{proof}

We also recall two standard consequences of 
rearrangements; see, e.g.~\cite[Chapter~3]{lieb-loss:book}. First, 
for a probability density \(g\) on $\RL^d$,
Riesz's rearrangement inequality implies that
\begin{equation*} 
  g^\star*g^\star
  \quad\text{majorizes}\quad g*g,
\end{equation*}
that is, for every measurable $E\subset\RL^d$ of finite 
Lebesgue measure,
\[
  \int_E(g*g)
  \le \int_{E^\star}(g^\star*g^\star).
\]
Second, for nonnegative functions $f,g$ and \(1\le p<\infty\),
\begin{equation}\label{eq:rearrange-contraction}
  \|f^\star-g^\star\|_p\le\|f-g\|_p.
\end{equation}

By passing to convergence of entropies of rearrangements, which will 
imply $L^2$ convergence by the lemma above, we will leverage a recent 
stability estimate in the sharp form of Young's 
inequality~\cite[Theorem~1.1]{christ:19}. 
We will also use the classical fact that 
every nonnegative nonzero extremizing pair
consists of Gaussian functions~\cite{brascamp:76}; 
see also~\cite[Proposition~3.1]{christ:19}.

\begin{lemma}\label{lem:young-gaussian}
Let \(\{g_n,\widehat g_n\}\) be probability
densities in \(L^{4/3}(\R^d)\), let \(\phi_{\Sigma}\) 
denote the centered Gaussian density
with covariance matrix \(\Sigma\), 
and let $\phi$ denote the standard normal density.
If, as $n\to\infty$,
\[
  \|g_n^\star-\phi\|_{4/3}
  ,\;\|\widehat g_n^\star-\phi\|_{4/3}
  \rightarrow0,
\]
and
\[
  \frac{\|g_n*\widehat g_n\|_2}
       {\|g_n\|_{4/3}\|\widehat g_n\|_{4/3}}
  \rightarrow
  \left(\frac{(4/3)^{3/4}}{\sqrt2}\right)^d,
\]
then there exist \(\{m_n,\widehat m_n\}\) in $\R^d$ and
positive definite matrices \(\{\Sigma_n\}\) with
\(\det(\Sigma_n)=1\), such that
\[
  \|g_n-\phi_{\Sigma_n}(\,\cdot-m_n)\|_1
,\;
  \|\widehat g_n
     -\phi_{\Sigma_n}(\,\cdot-\widehat m_n)\|_1
  \rightarrow0.
\]
\end{lemma}

\begin{proof}
Since rearrangement preserves the \(L^{4/3}\) norm,
\[
  \|g_n\|_{4/3},\ \|\widehat g_n\|_{4/3}
  \rightarrow\|\phi\|_{4/3}.
\]
Applying the stability theorem for Young's
inequality~\cite[Theorem~1.1]{christ:19}, we obtain Gaussian
extremizing functions \(\{F_n,\widehat F_n\}\) satisfying
\begin{equation} \label{approximationFg}
  \|g_n-F_n\|_{4/3}
  ,
  \|\widehat g_n-\widehat F_n\|_{4/3}
  \rightarrow0.
\end{equation}
We may take these functions to be nonnegative.
Indeed, replacing an extremizing pair by its absolute
values preserves extremality, since
\[
  \|F_n*\widehat F_n\|_2
  \leq\||F_n|*|\widehat F_n|\|_2
  \leq
  \left(\frac{(4/3)^{3/4}}{\sqrt2}\right)^d
  \|F_n\|_{4/3}\|\widehat F_n\|_{4/3},
\]
and does not increase the left-hand side in~\eqref{approximationFg}.
The equality characterization for the sharp form of Young's 
inequality~\cite{brascamp:76} implies that these nonnegative functions
are Gaussian and, since $p=q=4/3$,
they have the same quadratic form.

Using the rearrangement contraction 
inequality~\eqref{eq:rearrange-contraction} we get that,
\begin{align*}
  \|F_n^\star-\phi\|_{4/3}
  &\leq
  \|F_n^\star-g_n^\star\|_{4/3}
  +\|g_n^\star-\phi\|_{4/3}\\
  &\leq
  \|F_n-g_n\|_{4/3}
  +\|g_n^\star-\phi\|_{4/3}
  \rightarrow0,
\end{align*}
and the same holds for \(\widehat F_n^\star\).
We may write
\[
  F_n^\star(x)=\kappa_n e^{-\ell_n|x|^2}
  \quad \text{and} \quad
  \widehat F_n^\star(x)
  =\widehat\kappa_n e^{-\ell_n|x|^2}.
\]
We claim that
\begin{equation} \label{parameterclaim}
  \kappa_n,\ \widehat\kappa_n
  \rightarrow(2\pi)^{-d/2}
   \quad \text{and} \quad
  \ell_n\rightarrow\frac12.
\end{equation}
Every subsequence admits a further subsequence along which
both rearrangements converge almost everywhere to \(\phi\).
Choose two points \(x_1,x_2\) of convergence with
\(|x_1|\neq|x_2|\). Along this further subsequence,
\[
  \ell_n
  =
  \frac{
    \log F_n^\star(x_1)-\log F_n^\star(x_2)
  }{|x_2|^2-|x_1|^2}
  \rightarrow\frac12,
\]
and
$$
  \kappa_n
  =
  F_n^\star(x_1)e^{\ell_n|x_1|^2}
  \rightarrow(2\pi)^{-d/2}.
$$
Similarly,
$\widehat\kappa_n
  =
  \widehat F_n^\star(x_1)e^{\ell_n|x_1|^2}
  \rightarrow(2\pi)^{-d/2}.$
This proves the claim~\eqref{parameterclaim} along the full sequence.

The Gaussian functions need not have integral one, but
\[
  \int_{\R^d}F_n
  =\kappa_n\left(\frac{\pi}{\ell_n}\right)^{d/2}
  \rightarrow1
\]
and similarly for $\widehat F_n$. 
Since $F_n$ and $\widehat F_n$ have the same 
quadratic form, we can now write
$$
  F_n(x)
  =
  \kappa_n
  e^{-\ell_n\|\Sigma_n^{-1/2}(x-m_n)\|_2^2},
  $$
  and
  $$
  \widehat F_n(x)
  =
  \widehat\kappa_n
  e^{-\ell_n\|\Sigma_n^{-1/2}(x-\widehat m_n)\|_2^2},
$$
with 
 $ \Sigma_n\succ0$ and $\det(\Sigma_n)=1.$ 
Hence, by convergence of the parameters and dominated
convergence,
\[
  \|F_n(\Sigma_n^{1/2}\,\cdot+m_n)-\phi\|_{4/3}
  ,\;
  \|\widehat F_n(
      \Sigma_n^{1/2}\,\cdot+\widehat m_n)-\phi\|_{4/3}
  \rightarrow0.
\]
Indeed, for all sufficiently large \(n\), the two transformed
Gaussian functions are bounded by \(K e^{-|x|^2/4}\) for
some fixed \(K<\infty\).

Since \(\det(\Sigma_n^{1/2})=1\),~\eqref{approximationFg} now implies
\[
  \|g_n(\Sigma_n^{1/2}\,\cdot+m_n)-\phi\|_{4/3}
  ,\;
  \|\widehat g_n(
      \Sigma_n^{1/2}\,\cdot+\widehat m_n)-\phi\|_{4/3}
  \rightarrow0.
\]
Finally, 
by the preceding $L^{4/3}$ convergence every subsequence has a further 
subsequence converging to $\varphi$ a.e. Since the transformed densities 
and $\varphi$ all have integral one, Scheff\'{e}'s lemma gives 
$L^1$ convergence along this further subsequence, and hence 
along the whole sequence.
And
since \(\det(\Sigma_n)=1\) implies that
\[
  \phi_{\Sigma_n}(x-m_n)
  =\phi\bigl(\Sigma_n^{-1/2}(x-m_n)\bigr),
\]
the conclusion follows.
\end{proof}

\subsection{Total-variation and Ruzsa stability}

The proof of Theorem~\ref{prop:tv} starts with smoothing. 
We need the following technical lemma to ensure integrability of the 
logarithms
of the Jacobian determinants and diagonal entries. 
The notation $X \stackrel{\mathrm{d}}{=} Y$ denotes equality of 
distributions, i.e., that $X$ has the same distribution as $Y.$
\begin{lemma}\label{lem:triangular-transports}
Let \(X,Y\) be independent random vectors
in \(\R^d\), with densities \(f_X,f_Y\), respectively.
Suppose that each density is the convolution of a compactly
supported probability law with a nondegenerate Gaussian density.
Let \(\phi\) denote the standard Gaussian density on \(\R^d\).

There exist continuously differentiable triangular maps
\(T,\widehat T:\R^d\to\R^d\), whose Jacobian matrices, 
$DT$ and $D\widehat T$, respectively, have positive diagonal 
entries, such that
\[
  T(G)\stackrel{\mathrm{d}}{=}X
  \quad \text{and} \quad
  \widehat T(G')\stackrel{\mathrm{d}}{=}Y,
\]
where \(G,G'\) are independent standard Gaussian vectors.
Moreover,
\[
  f_X(T(u))\det (DT(u))=\phi(u)\quad \text{and} \quad
  f_Y(\widehat T(u))
  \det (D\widehat T(u))=\phi(u),
\]
hence,
\[
  \E|\log\det (DT(G))|,\;
  \E|\log\det (D\widehat T(G'))|<\infty
\]
and for all \(i=1,\ldots,d\),
\[
  \E|\log\partial_iT_i(G)|,\;
  \E|\log\partial_i\widehat T_i(G')|<\infty.
\]
\end{lemma}

\begin{proof}
The densities of \(X\) and \(Y\) are everywhere positive and
in \(C^{\infty}(\mathbb R^d)\). Therefore,~\cite[Lemma~1]{rioul:17} 
guarantees the existence of the maps
\(T,\widehat T\) from the standard Gaussian law to \(f_X,f_Y\),
respectively. From the same lemma these are continuously differentiable
and their Jacobian matrices have positive diagonal entries.

Write \(X^i=(X_1,\ldots,X_i)\) and
\(Y^i=(Y_1,\ldots,Y_i)\), with densities
\(f_{X^i}\) and \(f_{Y^i}\), respectively, and let
\[
  T^i(u)=\bigl(T_1(u),\ldots,T_i(u)\bigr)\quad \text{and} \quad
  \widehat T^i(u)
  =\bigl(\widehat T_1(u),\ldots,\widehat T_i(u)\bigr).
\]
Since \(T\) is triangular, \(T^i\) transports
the first \(i\) coordinates of the Gaussian to \(X^i\).
Thus, by a change of variables,
\begin{equation}\label{changeofvars}
  f_{X^i}\bigl(T^i(u)\bigr)
  \prod_{j=1}^i\partial_jT_j(u)
  =
  (2\pi)^{-i/2}
  \exp\left(-\frac12\sum_{j=1}^iu_j^2\right).
\end{equation}
For \(i=d\), this is the asserted identity,
\[
  f_X(T(u))\det (DT(u))=\phi(u),
\]
and it follows that
\[
\begin{aligned}
  \E|\log\det (DT(G))|
  &=
  \E\bigl|\log\phi(G)-\log f_X(T(G))\bigr|\\
  &\leq
  \E|\log\phi(G)|
  +\int_{\R^d}f_X(x)|\log f_X(x)|\,dx
  <\infty,
\end{aligned}
\]
where the last term is finite since \(f_X\) is assumed to be a
convolution of a compactly supported measure with the Gaussian.

By~\eqref{changeofvars},
\[
  \frac{
    f_{X^i}\bigl(T^i(u)\bigr)
  }{
    f_{X^{i-1}}\bigl(T^{i-1}(u)\bigr)
  }
  \,\partial_iT_i(u)
  =
  \frac1{\sqrt{2\pi}}e^{-u_i^2/2},
\]
where, for \(i=1\), the denominator is interpreted as \(1\).
Therefore,
\[
\begin{aligned}
  \log\partial_iT_i(G)
  =&
  -\frac12\log(2\pi)-\frac12G_i^2\\
  &-\log f_{X^i}(T^i(G))
  +\log f_{X^{i-1}}(T^{i-1}(G)),
\end{aligned}
\]
which implies that,
\[
\begin{aligned}
  \E|\log\partial_iT_i(G)|
  <\infty,
\end{aligned}
\]
by the integrability of $\log f_{X^i}$.

The same argument applies to \(\widehat T\), using the
densities \(f_{Y^i}\).
\end{proof}

We are finally ready to give the proof of Theorem~\ref{prop:tv}:
\begin{proof}[Proof of Theorem~\ref{prop:tv}]
By the EPI,
\begin{align*}
  \delta_{\mathrm{EPI}}(X_n,Y_n)
  &\ge
  \frac d2\log\!\left(
    \frac{e^{2h(X_n)/d}+e^{2h(Y_n)/d}}2
  \right)
  -\frac{h(X_n)+h(Y_n)}2\\
  &=
  \frac d2\log\cosh\!\left(
    \frac{h(X_n)-h(Y_n)}d
  \right).
\end{align*}
Since the deficit tends to zero and \(d\) is fixed,
we have that
$|h(X_n)-h(Y_n)|\rightarrow0$, 
as $n\to\infty$, which 
proves the first assertion.

By scaling each pair \(X_n,Y_n\) by the same positive number,
we may now assume without loss of generality that
\begin{equation} \label{scalingassumption}
  \frac{h(X_n)+h(Y_n)}2=\frac d2\log(2\pi e).
\end{equation}
It follows that, as $n\to\infty$,
\[
  h(X_n),\ h(Y_n)\rightarrow\frac d2\log(2\pi e),
  \]
i.e.,  $
  N(X_n),\ N(Y_n)\rightarrow1.
$
Our goal now is to show that
\[
  \inf_{\substack{
    m,\widehat m\in\R^d\\
    \Sigma\succ0,\ \det(\Sigma)=1}}
  \left(
    \|f_n-\phi_\Sigma(\,\cdot-m)\|_1
    +
    \|\widehat f_n-\phi_\Sigma(\,\cdot-\widehat m)\|_1
  \right)
  \rightarrow0.
\]

We first assume that every \(f_n,\widehat f_n\) is the convolution
of a compactly supported probability density with a nondegenerate
isotropic Gaussian density. In particular, these densities are
smooth and strictly positive.
Let \(T_n,\widehat T_n\) be the transport maps from Lemma~\ref{lem:triangular-transports} applied to $X_n$ and $Y_n$. 
Thus,  $T_n(G)$ and $\widehat T_n(G')$ have the same distributions 
as $X_n$ and $Y_n$ respectively, where $G,G'$ are 
independent standard normals. Also,
\[
  f_n(T_n(u))\det(DT_n(u))=\phi(u)
, \qquad
  \widehat f_n(\widehat T_n(u))
  \det (D\widehat T_n(u))=\phi(u),
\]
and
\begin{equation}\label{eq:transport-mean}
\begin{aligned}
  \E[\log\det( DT_n(G))]
  &=h(X_n)-h(G)\rightarrow0,\\
  \E[\log\det( D\widehat T_n(G'))]
  &=h(Y_n)-h(G')\rightarrow0.
\end{aligned}
\end{equation}
Note that the logarithms on the left-hand side are integrable, since
\[
  \log\det(DT_n(G))
  =\log\phi(G)-\log f_n(T_n(G)),
\]
and similarly for \(\widehat T_n\).

We use a transport-rotation argument similar to the one
in~\cite{rioul:17} for the EPI. Let \(G_1,G_2\) be independent
standard Gaussian vectors, and write
\[
  U=\frac{G_1+G_2}{\sqrt2}\quad \text{and} \quad
  V=\frac{G_1-G_2}{\sqrt2},
\]
so that the vectors \(U,V\) are independent standard Gaussians. 
Define
\[
  S_n(u,v)=\frac1{\sqrt2}\left[
    T_n\!\left(\frac{u+v}{\sqrt2}\right)
    +\widehat T_n\!\left(\frac{u-v}{\sqrt2}\right)
  \right],
  \qquad u,v\in\R^d.
\]
Then
\[
  S_n(U,V)
  =\frac{T_n(G_1)+\widehat T_n(G_2)}{\sqrt2},
\]
has the same law as \((X_n+Y_n)/\sqrt2\).
For each fixed \(v\in\R^d\),
\[
  D_uS_n(u,v)
  =
  \frac12\left[
    DT_n\!\left(\frac{u+v}{\sqrt2}\right)
    +D\widehat T_n\!\left(\frac{u-v}{\sqrt2}\right)
  \right],
\]
and
by change of variables
\[
  h(S_n(U,V),V)
  =
  h(U,V)+\E[\log\det(D_uS_n(U,V))].
\]
This is analogous to~\cite[Lemma~2]{rioul:17}. To justify it, note 
that the matrix $D_uS_n(u,v)$ is triangular with positive diagonal entries.
Thus, \(u\mapsto S_n(u,v)\) is continuously differentiable
and injective. The map \((u,v)\mapsto(S_n(u,v),v)\) is therefore continuously
differentiable and injective as well, with Jacobian 
\(\det(D_uS_n(u,v)) >0\).
Moreover, using Lemma~\ref{lem:triangular-transports},
$$
  \E\bigl|\log\det(D_uS_n(U,V))\bigr|
\le
  \sum_{i=1}^d
  \left(
    \E|\log\partial_iT_{n,i}(G_1)|
    +
    \E|\log\partial_i\widehat T_{n,i}(G_2)|
  \right)\\
<\infty.
$$

Since \(h(S_n(U,V))\) is finite by assumption
and since $U,V$ are independent, we obtain that
\begin{align*}
  h\left(\frac{X_n+Y_n}{\sqrt2}\right)
  &=h(S_n(U,V))\\
  &\ge h(S_n(U,V)|V)\\
  &=h(S_n(U,V),V)-h(V)\\
  &=h(U,V)-h(V)+\E[\log\det(D_uS_n(U,V))]\\
  &=h(G)+
    \E\left[\log\det\left(
      \frac{DT_n(G_1)+D\widehat T_n(G_2)}2
    \right)\right].
\end{align*}
Therefore, using the triangularity of the transport maps and  
the simple identity,
$$  \log\frac{a+b}{2}
  =
  \frac{\log a+\log b}{2}
  +\log\cosh\!\left(\frac{\log a-\log b}{2}\right),
  \quad a,b>0,
$$ 
we have  
\begin{align*}
  \delta_{\mathrm{EPI}}(X_n,Y_n)
  &\geq \E\left[\log\det\!\left(
    \frac{DT_n(G_1)+D\widehat T_n(G_2)}2
  \right)\right]\\
  &= \E\left[\sum_{i=1}^d
  \log\left(
    \frac{
      \partial_iT_{n,i}(G_1)
      +\partial_i\widehat T_{n,i}(G_2)
    }2
  \right)\right],
\end{align*}
and hence,
\begin{align}
  \delta_{\mathrm{EPI}}(X_n,Y_n)
  &\geq \frac12\E[\log\det(DT_n(G_1))]
     +\frac12\E[\log\det(D\widehat T_n(G_2))]
     \notag\\
  &\qquad+
  \E\left[\sum_{i=1}^d
  \log\cosh\left(
    \frac{
      \log\partial_iT_{n,i}(G_1)
      -\log\partial_i\widehat T_{n,i}(G_2)
    }2
  \right)\right]\notag \\
  &=
  \E\left[\sum_{i=1}^d
  \log\cosh\!\left(
    \frac{
      \log\partial_iT_{n,i}(G_1)
      -\log\partial_i\widehat T_{n,i}(G_2)
    }2
  \right)\right],
	\label{lastcosh}
  \end{align}
where the last equality follows by~\eqref{eq:transport-mean} and the 
scaling assumption~\eqref{scalingassumption}. Using the convexity 
of $x \mapsto \log \cosh x$, it follows from~\eqref{lastcosh} that
$$
 \delta_{\mathrm{EPI}}(X_n,Y_n) 
  \geq
  d\,\E\left[\log\cosh\left(
    \frac1d\sum_{i=1}^d
    \frac{
      \log\partial_iT_{n,i}(G_1)
      -\log\partial_i\widehat T_{n,i}(G_2)
    }2
  \right)\right],
$$
  and using the triangular structure of the transport maps again, 
  \begin{align*}
 \delta_{\mathrm{EPI}}(X_n,Y_n)&\geq 
  d\,\E\left[\log\cosh\left(
    \frac{
      \log\det(DT_n(G_1))
      -\log\det(D\widehat T_n(G_2))
    }{2d}
  \right)\right]\\
  &=
  d\,\E\left\{\E\Biggl[
    \log\cosh\!\left(
      \frac{
        \log\det(DT_n(G_1))
        -\log\det(D\widehat T_n(G_2))
      }{2d}
    \right)
    \,\Bigg|\,G_1
  \Biggr]
	\right\}\\
  &\geq
  d\,\E\left[\log\cosh\left(
    \frac{
      \log\det(DT_n(G_1))
      -\E\bigl[\log\det(D\widehat T_n(G_2))\big| G_1\bigr]
    }{2d}
  \right)\right]\\
  &=
  d\,\E\left[\log\cosh\left(
    \frac{
      \log\det(DT_n(G))
      -\E\log\det(D\widehat T_n(G))
    }{2d}
  \right)\right].
\end{align*}
The same bound holds after interchanging \(T_n\) and
\(\widehat T_n\). Since both expectations
in~\eqref{eq:transport-mean} tend to zero 
and $\cosh(x) \geq 1$ for all $x\in \mathbb{R}$, it follows that
\[
  \log\det(DT_n(G))\rightarrow0\quad \text{and} \quad
  \log\det(D\widehat T_n(G))\rightarrow0 \quad \text{in probability.}
\]

Now choose a nonnegative sequence
\(\{\eta_n\}\) converging to zero, such that, if we write
\[
\begin{aligned}
  E_n&=\{u:|\log\det(DT_n(u))|\leq\eta_n\},
  &p_n&=\BBP (G\in E_n), \\
  \widehat E_n
  &=\{u:|\log\det(D\widehat T_n(u))|\leq\eta_n\}, 
  &\widehat p_n&=\BBP (G\in\widehat E_n),
\end{aligned}
\]
then \(p_n,\widehat p_n\to1\).
We may assume without loss of generality that
\(p_n,\widehat p_n>0\).
Let \(g_n\) be the density of \(T_n(G)\) conditional on
\(G\in E_n\), and let \(\widehat g_n\) be the density of
\(\widehat T_n(G)\) conditional on \(G\in\widehat E_n\).
For \(p_n<1\) and \(\widehat p_n<1\), let
\(b_n,\widehat b_n\) be the corresponding conditional
densities on the complements. Then we write
\begin{equation}\label{eq:truncation}
\begin{aligned}
  f_n&=p_ng_n+(1-p_n)b_n,\\
  \widehat f_n
  &=\widehat p_n\widehat g_n
    +(1-\widehat p_n)\widehat b_n.
\end{aligned}
\end{equation}
Since $h(f_n)$ is finite, $h(g_n), h(b_n)$ are finite as well. 

For every \(t>0\), a change of variables yields
\begin{align*}
  |\{g_n>t\}|
  &=
  \int_{E_n}
    \IND\left\{
      \frac{\phi(u)}{p_n\det(DT_n(u))}>t
    \right\}
    \det(DT_n(u))\,du\\
  &\le
  e^{\eta_n}
  |\{\phi>p_ne^{-\eta_n}t\}|.
\end{align*}
The same argument applies to \(\widehat g_n\).
It follows that
\begin{equation}\label{eq:gaussian-envelope}
\begin{aligned}
  g_n^\star(x)
  &\le
  \frac{e^{\eta_n}}{p_n}
  \phi(e^{-\eta_n/d}x),\\
  \widehat g_n^\star(x)
  &\le
  \frac{e^{\eta_n}}{\widehat p_n}
  \phi(e^{-\eta_n/d}x).
\end{aligned}
\end{equation}
Both right-hand sides converge to \(\phi\) in \(L^1\),
and
their integrals, \(e^{2\eta_n}/p_n\) and
\(e^{2\eta_n}/\widehat p_n\), respectively, also tend to one.
Since \(g_n^\star,\widehat g_n^\star\) have integral one,
they converge to \(\phi\) in \(L^1\).
And since the right-hand side is uniformly bounded, we conclude
that
\begin{equation}\label{eq:rearranged-convergence}
  g_n^\star,\ \widehat g_n^\star
  \rightarrow\phi
  \quad\text{in }L^p
  \text{ for every }1\leq p<\infty.
\end{equation}
We will use~\eqref{eq:rearranged-convergence} for \(p=4/3\)
in order to use the stability of Young's inequality.

For all sufficiently large \(n\),~\eqref{eq:gaussian-envelope} yields
\[
  g_n^\star(x),\ \widehat g_n^\star(x)
  \leq K e^{-c|x|^2},
\]
for some fixed \(K,c>0\).
In particular, these imply that 
\(|g_n^\star\log g_n^\star|\) and
\(|\widehat g_n^\star\log\widehat g_n^\star|\) are dominated 
on a bounded range, since \(|t\log t|\leq K'\sqrt t\).
Every subsequence of~\eqref{eq:rearranged-convergence}
has an almost-everywhere convergent subsequence, so by
dominated convergence 
\begin{equation}\label{eq:good-entropy}
\begin{aligned}
  h(g_n)=h(g_n^\star)
  &\rightarrow\frac d2\log(2\pi e), \\
  h(\widehat g_n)=h(\widehat g_n^\star)
  &\rightarrow\frac d2\log(2\pi e).
\end{aligned}
\end{equation}
Moreover, 
\begin{align}
  (1-p_n)h(b_n)
  &=h(X_n)-p_nh(g_n)+p_n\log p_n 
     +(1-p_n)\log(1-p_n)
  \rightarrow0, \notag\\
  (1-\widehat p_n)h(\widehat b_n)
  &=h(Y_n)-\widehat p_nh(\widehat g_n)
      +\widehat p_n\log\widehat p_n 
     +(1-\widehat p_n)\log(1-\widehat p_n)
  \rightarrow0.
  \label{eq:bad-entropy}
\end{align}

We claim that
\begin{equation}\label{eq:convolution-entropies}
  h(g_n*\widehat g_n),\
  h(g_n^\star*\widehat g_n^\star)
  \rightarrow
  \frac d2\log(2\pi e)+\frac d2\log2.
\end{equation}
For \(g_n*\widehat g_n\), the lower bound follows from
the EPI and~\eqref{eq:good-entropy}.
For the upper bound, 
we observe that by~\eqref{eq:truncation}, concavity of entropy and the EPI,
\ben
  h(f_n*\widehat f_n)
  &\ge&
  p_n\widehat p_n h(g_n*\widehat g_n)
  +p_n(1-\widehat p_n)h(g_n*\widehat b_n)
  +(1-p_n)\widehat p_n h(b_n*\widehat g_n)
  \notag\\
  &&\;+(1-p_n)(1-\widehat p_n)h(b_n*\widehat b_n)
  \notag\\
  &\ge&
  p_n\widehat p_n h(g_n*\widehat g_n)
  +\frac{p_n(1-\widehat p_n)}2h(g_n)
  +\frac{(1-p_n)\widehat p_n}2h(\widehat g_n)
 +\frac{1-p_n}2h(b_n)
  \notag\\
  &&\; +\frac{1-\widehat p_n}2h(\widehat b_n)
  +(1-p_n\widehat p_n)\frac d2\log2.
  \label{eq:convolution-truncation}
\een
The left-hand side converges to
\(\frac d2\log(2\pi e)+\frac d2\log2\),
and~\eqref{eq:good-entropy} and~\eqref{eq:bad-entropy} 
give the upper bound on
\(h(g_n*\widehat g_n)\).

For \(g_n^\star*\widehat g_n^\star\), writing
\[
  g_n^\star*\widehat g_n^\star-\phi*\phi
  =
  (g_n^\star-\phi)*\widehat g_n^\star
  +\phi*(\widehat g_n^\star-\phi),
\]
and using elementary convolution bounds we obtain
\begin{align*}
  \|g_n^\star*\widehat g_n^\star-\phi*\phi\|_1
  &\le
  \|g_n^\star-\phi\|_1
  +\|\widehat g_n^\star-\phi\|_1
  \rightarrow0,\\
\|g_n^\star*\widehat g_n^\star-\phi*\phi\|_\infty
  &\le
  \|\widehat g_n^\star\|_\infty
  \|g_n^\star-\phi\|_1+
  \|\phi\|_\infty
  \|\widehat g_n^\star-\phi\|_1
  \rightarrow0,
\end{align*}
where we used~\eqref{eq:rearranged-convergence}, 
the fact that 
$g_n^\star$ and $\phi$ integrate to one,
and the uniform bound on
\(\|g_n^\star\|_\infty\) from~\eqref{eq:gaussian-envelope}.
Hence,
\begin{equation}\label{L2conv}
  \|g_n^\star*\widehat g_n^\star-\phi*\phi\|_2^2
  \le
  \|g_n^\star*\widehat g_n^\star-\phi*\phi\|_\infty\cdot
  \|g_n^\star*\widehat g_n^\star-\phi*\phi\|_1
  \rightarrow0.
\end{equation}

To justify convergence of the entropies, choose fixed
\(K,c>0\) such that
\(g_n^\star(x),\widehat g_n^\star(x)\leq K e^{-c|x|^2}\)
for all sufficiently large \(n\), as permitted 
by~\eqref{eq:gaussian-envelope}. Then
\[
  0\leq(g_n^\star*\widehat g_n^\star)(x)
  \leq
  K^2\int_{\R^d}e^{-c|y|^2-c|x-y|^2}\,dy
  =
  K^2\left(\frac{\pi}{2c}\right)^{d/2}
  e^{-c|x|^2/2}.
\]
Since \(|t\log t|\leq C\sqrt t\) on any fixed bounded
interval of nonnegative \(t\), it follows that
\[
  \left|
    (g_n^\star*\widehat g_n^\star)(x)
    \log(g_n^\star*\widehat g_n^\star)(x)
  \right|
  \leq C e^{-c|x|^2/4},
\]
where \(C\) is independent of \(n\).
The right-hand side is integrable,
and 
the uniform convergence established above gives
\[
  (g_n^\star*\widehat g_n^\star)(x)
  \rightarrow(\phi*\phi)(x),
\]
so by dominated convergence 
\[
  h(g_n^\star*\widehat g_n^\star)
  \rightarrow h(\phi*\phi)
  =
  \frac d2\log(2\pi e)+\frac d2\log2.
\]
This proves the claim for
\(g_n^\star*\widehat g_n^\star\) 
in~\eqref{eq:convolution-entropies}.

Note that 
\[
  \|g_n*\widehat g_n\|_\infty,\
  \|g_n^\star*\widehat g_n^\star\|_\infty
  \leq
  \|g_n\|_\infty
  =
  \|g_n^\star\|_\infty.
\]
By Riesz's rearrangement inequality,
\(g_n^\star*\widehat g_n^\star\) majorizes
\(g_n*\widehat g_n\),
so
Lemma~\ref{lem:majorization} and~\eqref{eq:convolution-entropies}
imply
\[
  \|g_n^\star*\widehat g_n^\star\|_2^2
  -
  \|g_n*\widehat g_n\|_2^2
  \rightarrow0.
\]
By~\eqref{L2conv} we know that
\(g_n^\star*\widehat g_n^\star\to\phi*\phi\) in \(L^2\),
and by~\eqref{eq:rearranged-convergence} 
with \(p=4/3\) and
preservation of \(L^p\) norms under rearrangement
we also have that
\[
  \|g_n\|_{4/3},\ \|\widehat g_n\|_{4/3}
  \rightarrow\|\phi\|_{4/3}.
\]
Therefore,
\[
  \frac{\|g_n*\widehat g_n\|_2}
       {\|g_n\|_{4/3}\|\widehat g_n\|_{4/3}}
  \rightarrow
  \frac{\|\phi*\phi\|_2}{\|\phi\|_{4/3}^2}
  =
  \left(\frac{(4/3)^{3/4}}{\sqrt2}\right)^d.
\]
By Lemma~\ref{lem:young-gaussian} there exist $m_n,\widehat{m}_n \in \mathbb R^d$ and positive definite matrices $\Sigma_n$ with $\mathrm{det}(\Sigma_n) = 1$ such that 
\[
  \|g_n-\phi_{\Sigma_n}(\,\cdot-m_n)\|_1
,
  \|\widehat g_n
     -\phi_{\Sigma_n}(\,\cdot-\widehat m_n)\|_1
  \rightarrow0.
\]
Finally,
\[
  \|f_n-g_n\|_1=2(1-p_n)\rightarrow0
  \quad \text{and} \quad
  \|\widehat f_n-\widehat g_n\|_1
  =2(1-\widehat p_n)\rightarrow0.
\]
Since \(N(X_n),N(Y_n)\to1\) and the
\(L^1\) distance is affine invariant, we also have
\[
  \|\phi_{N(X_n)\Sigma_n}-\phi_{\Sigma_n}\|_1
  =
  \|\phi_{N(X_n)I_d}-\phi\|_1
  \rightarrow0,
\]
and similarly for \(Y_n\).
This proves the theorem
under the additional
regularity assumption
that each $f_n$ and $\widehat f_n$
is the convolution of a compactly supported probability
density with a nondegenerate isotropic Gaussian density.

To prove the general case, consider first the truncated densities
\[
  f_{n,R}(x)
  =
  \frac{f_n(x)\IND\{|x|\leq R\}}
       {\int_{|y|\leq R}f_n(y)\,dy}\quad \text{and} \quad
  \widehat f_{n,R}(x)
  =
  \frac{\widehat f_n(x)\IND\{|x|\leq R\}}
       {\int_{|y|\leq R}\widehat f_n(y)\,dy},
	\quad x\in\RL^d,
\]
for all sufficiently large \(R\).
These densities converge to \(f_n,\widehat f_n\) in
\(L^1\), and by a direct application of dominated convergence 
\[
  h(f_{n,R})\rightarrow h(f_n) \quad
  \text{and} \quad
  h(\widehat f_{n,R})\rightarrow h(\widehat f_n) \quad \text{as } R\to \infty.
\]
Furthermore,
\[
  (f_n\IND\{|\cdot|\leq R\})
  *
  (\widehat f_n\IND\{|\cdot|\leq R\})
  \uparrow f_n*\widehat f_n
  \quad\text{a.e.},
\]
and another application of dominated convergence yields
\[
  h(f_{n,R}*\widehat f_{n,R})
  \rightarrow h(f_n*\widehat f_n).
\]
By Lemma~\ref{lem:regularization}, for each \(n\), we may first choose \(R_n\)
and then \(0<t_n<1/n\) such that the independent random
vectors \(\widetilde X_n,\widetilde Y_n\), with densities
$
  f_{n,R_n}*\phi_{t_nI_d}
$ and $
  \widehat f_{n,R_n}*\phi_{t_nI_d}
$ respectively,
satisfy
\begin{align*}
  \TV(\widetilde X_n,X_n)
  ,\TV(\widetilde Y_n,Y_n)
  &<1/n,\\
  |h(\widetilde X_n)-h(X_n)|
  ,|h(\widetilde Y_n)-h(Y_n)|
  &<1/n,\\
   |h(\widetilde X_n+\widetilde Y_n)-h(X_n+Y_n)|
  &<1/n,
\end{align*}
so that
\(\delta_{\mathrm{EPI}}(\widetilde X_n,\widetilde Y_n)\to0\).
Letting
\[
  \lambda_n
  =
  \exp\!\left(
    \frac{h(\widetilde X_n)-h(X_n)}d
  \right),
	\qquad
  \widehat\lambda_n
  =
  \exp\!\left(
    \frac{h(\widetilde Y_n)-h(Y_n)}d
  \right),
\]
we have that \(\lambda_n,\widehat\lambda_n\to1\),
and 
applying the result already proved to
\(\widetilde X_n,\widetilde Y_n\),
we obtain
\(m_n,\widehat m_n\in\R^d\) and
\(\Sigma_n\succ0\), with \(\det(\Sigma_n)=1\), such that
(with a slight abuse of the notation for $\mathrm{TV}$-distance), 
\[
\begin{aligned}
  &\TV\!\left(
    \widetilde X_n,
    \mathcal{N}\big(m_n,N(\widetilde X_n)\Sigma_n\big)
  \right),
  \TV\!\left(
    \widetilde Y_n,
    \mathcal{N}\big(\widehat m_n,N(\widetilde Y_n)\Sigma_n\big)
  \right)
  \rightarrow0.
\end{aligned}
\]
Using the triangle inequality and the affine invariance of total
variation, we conclude that
\begin{align*}
  \TV\!\left(
    X_n,\mathcal{N}\big(m_n,N(X_n)\Sigma_n\big)
  \right)\leq
  \TV(X_n,\widetilde X_n)
  &+
  \TV\!\left(
    \widetilde X_n,
    \mathcal{N}\big(m_n,N(\widetilde X_n)\Sigma_n\big)
  \right)\\
  & +
  \TV\!\left(\mathcal{N}(0,\lambda_n^2I_d),\mathcal{N}(0,I_d)\right)
  \rightarrow0.
\end{align*}
The same argument, with $\widehat{\lambda}_n$, 
yields
$$
  \TV\left(
    Y_n,\mathcal{N}\big(\widehat m_n,N(Y_n)\Sigma_n\big)
  \right)
  \rightarrow0,
$$
with the same matrix $\Sigma_n$. By the first assertion of the theorem,
$$
\frac{N(X_n)}{h_n}
=\exp\left(\frac{h(X_n)-h(Y_n)}{d}\right)\to 1,\quad\mbox{and}\quad
\frac{N(Y_n)}{h_n}
=\exp\left(\frac{h(Y_n)-h(X_n)}{d}\right)\to 1.
$$
For any \(a>0\) and \(m\in\mathbb R^d\), applying the affine transformation \(x\mapsto h_n^{-1/2}\Sigma_n^{-1/2}(x-m)\) to both Gaussian laws gives
\[
\operatorname{TV}\left(\clN(m,a\Sigma_n),\clN(m,h_n\Sigma_n)\right)
=
\operatorname{TV}\left(\clN(0,(a/h_n)I_d),\clN(0,I_d)\right).
\]
Taking \(a=N(X_n)\) or \(a=N(Y_n)\), the right-hand side tends to zero 
in both cases, since the corresponding Gaussian densities converge 
to the standard Gaussian density and hence in $L^1$ by Scheff\'{e}'s lemma. 
The triangle inequality now gives,
\[
\|f_n-\phi_{m_n,h_n\Sigma_n}\|_1
+
\|\widehat f_n-\phi_{\widehat m_n,h_n\Sigma_n}\|_1
\to 0,
\]
and since \(\det(\Sigma_n)=1\), this proves the result.
\end{proof}

For the proof of the EPI stability in Ruzsa distance
in Corollary~\ref{corollaryruzsa} we will need the following
lemma. Recall the definition of the Ruzsa distance
from~\eqref{eq:RuzsaC} in the Introduction.

\begin{lemma}\label{lem:overlap}
Let \(X,X'\) be i.i.d.\ \(\mathbb R^d\)-valued random vectors with finite
entropy $h(X)=h(X')$, such that \(h(X+X')<\infty\).
Let \(Z\) be a Gaussian with the same entropy as \(X\)
and suppose that the densities $f_X,f_Z$ of \(X,Z\) are mixtures 
with a common component, that is, they may be expressed as
\[
  f_X=pw+(1-p)v
  \quad\text{and}\quad
  f_Z=pw+(1-p)u
\]
for some \(p\in(0,1)\) and probability densities \(w,v,u\) on
\(\mathbb R^d\). Then
\begin{align}
  d_R(X,Z)\le&
  \frac{d_R(X,-X')}{p}
  +H_2(p) +(1-p)\left(
    \frac d2\log\frac1{p(1-p)}+\frac d2-\log p
  \right),
\label{eq:overlap}
\end{align}
where $H_2(p)=-p\log p-(1-p)\log(1-p)$, $p\in(0,1)$, denotes
the binary entropy function.
\end{lemma}

\begin{proof}
Let \(W,V,U\) be independent random vectors, also independent of \(X\), with
densities \(w,v,u\), respectively.
Note that \(h(w),h(v)\) and \(h(u)\) are finite, and so are
\(h(X+W)\) and \(h(X+V)\).

Using the concavity of entropy, we have
\begin{align*}
  h(X+X')
  &\ge p\,h(X+W)+(1-p)h(X+V)\\
  &\ge p\,h(X+W)+(1-p)h(X),
\end{align*}
and thus,
\begin{equation}\label{eq:component-bound}
  h(X+W)-h(X)\le \frac{d_R(X,-X')}{p}.
\end{equation}
By the Ruzsa triangle inequality for differential entropy applied
to \(X\), \(-W\), and \(-U\),
\begin{equation}\label{triangleXUW}
  h(X+U)\le h(X+W)+h(W-U)-h(W).
\end{equation}
Recall the following elementary inequality
for \(f=pf_1+(1-p)f_2\), 
\[
  h(f)\le p h(f_1)+(1-p)h(f_2)+H_2(p).
\]
Applying this to \(X+Z\), taking $Z$ to be independent
of $X$, and noticing that
\(h(X-Z)=h(X+Z)\), we obtain
\[
  d_R(X,Z)=h(X+Z)-h(X)
  \le p h(X+W)+(1-p)h(X+U)+H_2(p)-h(X).
\]
Thus, using~\eqref{eq:component-bound} and~\eqref{triangleXUW},
\begin{align}
  d_R(X,Z)
  \le&
  \frac{d_R(X,-X')}{p}
  +H_2(p) +(1-p)\bigl(h(W-U)-h(W)\bigr).
  \label{eq:overlap-intermediate}
\end{align}

The pointwise inequality \(pw\le f_Z\) and
the fact that
\(-\log\|f_Z\|_\infty=h(Z)-\frac d2\) imply
\begin{equation}\label{hWlb}
  h(W)\ge -\log\|w\|_\infty
  \ge h(Z)+\log p-\frac d2.
\end{equation}
Writing \(m=\mathbb E Z\), \(\Sigma=\operatorname{Cov}(Z)\) and
using the bounds \(pw\le f_Z\) and \((1-p)u\le f_Z\), we also have
\[
  \mathbb E\bigl[(W-m)(W-m)^\top\bigr]\preceq\frac{\Sigma}{p}
  \quad\text{and}\quad
  \mathbb E\bigl[(U-m)(U-m)^\top\bigr]\preceq\frac{\Sigma}{1-p}.
\]
Therefore, since \(W\) and \(U\) are independent by definition,
\[
  \operatorname{Cov}(W-U)
  =\operatorname{Cov}(W)+\operatorname{Cov}(U)
  \preceq \frac{\Sigma}{p(1-p)}.
\]
Since the Gaussian maximizes entropy under fixed covariance,
\[
  h(W-U)
  \le \frac12\log\left(
    (2\pi e)^d\det\left(\frac{\Sigma}{p(1-p)}\right)
  \right)
  =h(Z)+\frac d2\log\frac1{p(1-p)}.
\]
Hence, by~\eqref{hWlb},
\[
  h(W-U)-h(W)
  \le \frac d2\log\frac1{p(1-p)}+\frac d2-\log p,
\]
and substituting into~\eqref{eq:overlap-intermediate} 
proves~\eqref{eq:overlap} as claimed.
\end{proof}

\begin{corollary}\label{corollaryruzsa}
 Let \(\{X_n\}\) be \(\mathbb R^d\)-valued random vectors
with finite entropies and such that
\[
  \delta_{\mathrm{EPI}}(X_n)\rightarrow0.
\]
Then there exist Gaussian random vectors $\{Z_n\}$ 
with \(h(Z_n)=h(X_n)\) for each $n$, such that
\[
  d_R(X_n,Z_n)\rightarrow\frac d2\log 2,
\quad\mbox{as}\;n\to\infty.
\]
\end{corollary}

\begin{proof}[Proof of Corollary~\ref{corollaryruzsa}]
By scaling, we may assume that
$h(X_n)=\frac d2\log(2\pi e)$ for all $n$.

Theorem~\ref{prop:tv} supplies Gaussian $Z_n$ with densities
\(\phi_{\Sigma_n}(\,\cdot-m_n)\), where
\(\Sigma_n\) is positive definite and \(\det(\Sigma_n)=1\), such that
\[
  \TV\bigl(X_n,Z_n\bigr)\rightarrow0.
\]

Writing \(f_n\) for the density of each \(X_n\), we have
\[
  \int_{\R^d}\min\{f_n(x),\phi_{\Sigma_n}(x-m_n)\}\,dx
  =1-\TV\bigl(X_n,\mathcal{N}(m_n,\Sigma_n)\bigr)\rightarrow1.
\]
Now we define
$$
  p_n=\frac{n}{n+1}
      \int_{\R^d}\min\{f_n(x),\phi_{\Sigma_n}(x-m_n)\}\,dx,
$$ 
and 
      $$
  w_n(x)=
  \frac{\min\{f_n(x),\phi_{\Sigma_n}(x-m_n)\}}
       {\displaystyle
        \int_{\R^d}\min\{f_n(y),\phi_{\Sigma_n}(y-m_n)\}\,dy}.
$$
Then \(0<p_n<1\), \(p_n\to1\), and \(w_n\) is a probability density.
Moreover, $p_nw_n(x)$ is bounded above by both 
$f_n(x)$ and $\phi_{\Sigma_n}(x-m_n)$.
Consequently the functions
\[
  v_n(x)=\frac{f_n(x)-p_nw_n(x)}{1-p_n}
  \quad \text{and} \quad
  u_n(x)=\frac{\phi_{\Sigma_n}(x-m_n)-p_nw_n(x)}{1-p_n}
\]
are nonnegative and have integral one. Thus, we may write
\[
  f_n=p_nw_n+(1-p_n)v_n
  \quad \text{and} \quad
  \phi_{\Sigma_n}(\,\cdot-m_n)=p_nw_n+(1-p_n)u_n,
\]
and we can apply Lemma~\ref{lem:overlap}.
Since
\[
  d_R(X_n,-X_n')\rightarrow\frac d2\log2
  \quad\text{and}\quad
  (1-p_n)|\log(1-p_n)|\rightarrow0,
\]
the right-hand side of~\eqref{eq:overlap} tends to
\(\frac d2\log2\). On the other hand, by the EPI we also have
\[
  d_R\bigl(X_n,Z_n\bigr)\ge\frac d2\log2,
\]
since $h(Z_n) = h(X_n)$, and thus we have 
established the desired convergence.
\end{proof}

Corollary~\ref{cor:bound} is a straightforward consequence of Corollary~\ref{corollaryruzsa}.

\section*{Acknowledgments}
L.G. and I.K. were supported in part by the {\small EPSRC}-funded 
{\small INFORMED-AI} project EP/Y028732/1. L.G. was supported in part 
by a grant of access to OpenAI models through the ChatGPT for Academic 
Researchers program. 
During the preparation of this manuscript, the authors used ChatGPT-6
for language editing and assistance in exploring 
and checking proof ideas in connection with
Theorem~\ref{prop:tv}. The authors provided the mathematical 
ideas and directed the development of the arguments; all results and 
proofs were written and independently verified by the authors, 
who take full responsibility for the content of the article.

\bibliographystyle{plain}
\bibliography{R_stability.bib}

\end{document}